\documentclass[aps,prl,twocolumn,showpacs,superscriptaddress,groupedaddress,amsmath,amssymb]{revtex4}
\usepackage{txfonts}
\usepackage{amsfonts}
\usepackage{setspace}
\usepackage{graphicx}     
\usepackage{bm}
\usepackage{float}
\usepackage{pstricks}
\usepackage{pst-node}
\usepackage{pst-blur}
\definecolor{Pink}{rgb}{1.,0.75,0.8}
\begin{document}

\title{Loop Current and Orbital Magnetism in Pyrite-Type $Ir_{3}Ch_{8}$ ($Ch=Se,Te$)}

\author{Xiaotian Zhang}
\affiliation{Beijing National Laboratory for Condensed Matter Physics, Institute of Physics, Chinese Academy of Sciences, Beijing 100190, China}

\author{Xianxin Wu}
\affiliation{Beijing National Laboratory for Condensed Matter Physics, Institute of Physics, Chinese Academy of Sciences, Beijing 100190, China}

\author{Di Xiao}
\affiliation{Department of Physics, Carnegie Mellon University, Pittsburg, PA 15213, USA}


\author{Jiangping Hu}
\affiliation{Beijing National Laboratory for Condensed Matter Physics, Institute of Physics, Chinese Academy of Sciences, Beijing 100190, China}\affiliation{Department of Physics, Purdue University, West Lafayette, Indiana 47907, USA}

\begin{abstract}
We study the parent compound of the pyrite-type $Ir_{x}Ch_{2}$ iridium chalcogenide superconductors, $Ir_{0.75}Ch_{2}$ (or $Ir_{3}Ch_{8}$).  While the lattice structure of the material is rather complicated, we show that the electron physics near the Fermi surface (FS) can  be described by a three-band tight-binding model. We find that the material  possesses a loop current ground state  which is responsible for the highly anisotropic strong diamagnetism observed in experiments. The fluctuations of this loop current state can be a possible pairing mechanism for superconductivity.
\end{abstract}

\date{\today}
\maketitle

{\it Introduction.}  Superconductivity in most unconventional superconductors discovered in the past decades  is close to an magnetically ordered state involving the spin degree of freedom. The effective magnetic interaction is created by the hopping between the $d-$ or $f-$ orbitals bridged by other orbitals (mostly $p-$orbitals) away from the Fermi surface\cite{Scalapino} and is believed to be responsible for  developing unconventional superconductivity\cite{Anderson}. The recently discovered superconducting Iridium compounds~\cite{Pyon, Yang, Hosono, Fang, Ootsuki, Zhou, Kis} are very intriguing new unconventional superconductors. The pyrite-type compounds exhibit a maximum $T_c=6.4K$ in $Ir_{0.91}Se_{2}$ and $4.7K$ in $Ir_{0.93}Te_{2}$, which are higher than in the $CdI_2$-type compounds~\cite{Pyon, Yang, Hosono}. The parent compound of pyrite-type $Ir_{x}Ch_{8}$ is identified as $Ir_{3}Ch_{8}$~\cite{Hosono}, which has an unit cell that is cubic above the structural transition temperature $T_{s}$ and rhombohedral below that~\cite{Hosono,Cao}. However, different from  other unconventional superconductors,  the new superconductor shows strong and highly anisotropic diamagnetism and a very weak magnetic moment which contributes to paramagnetism\cite{Cao}. The absence of strong spin magnetism is consistent with the fact that the $\sigma^{*}$ state composed by the $p-$orbital of the $Ch_{2}$ located at the center of the cubic unit cell has a considerable weight on the FS, which limits the development of spin interaction through super-exchange mechanism.  

 In this Letter, we investigate the origin of the diamagnetism and its relationship to superconductivity.  We find that the electronic physics near the Fermi surface can be described by a three-band tight-binding model, which is similar to other unconventional superconductors~\cite{Zhang,Kuroki,Graser,Jiangping}. However, unlike other unconventional superconductors the parent compounds of which have their magnetic properties dominated by spin antiferromagnetism, the high weight of the $Ch_{2}$ $p-$orbital on the FS limits the development of super-exchange antiferromagnetic coupling and the effect of on-site repulsive interactions. These materials can have a ground state with a loop electronic current, and hence an orbital magnetic moment, induced by the inter-site repulsion between the $Ir$ electrons. The results explain the experimentally observed strong and highly anisotropic diamagnetism~\cite{Cao}.  We also suggest that the fluctuations of loop currents can be a possible pairing mechanism as firstly proposed in cuprates~\cite{Varma2,Varma1}. 
\begin{figure}[htb]
\begin{center}
\includegraphics[width=0.6\linewidth]{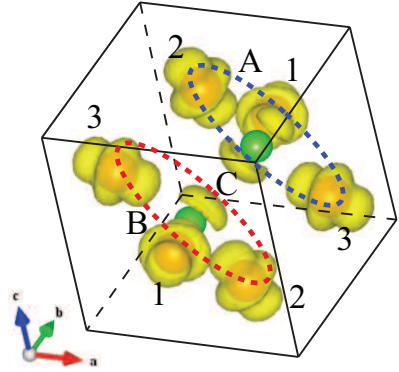}
\caption{The unit cell considered in the tight-binding model. The $Ir$ atoms are denoted as orange spheres while the $Ch_{2}$ is green. Isosurface of the electrons on the FS obtained by DFT calculations is plotted in yellow.}\label{structure}
\end{center}
\end{figure} 

\begin{figure}[htb]
\begin{center}
\includegraphics[width=70mm]{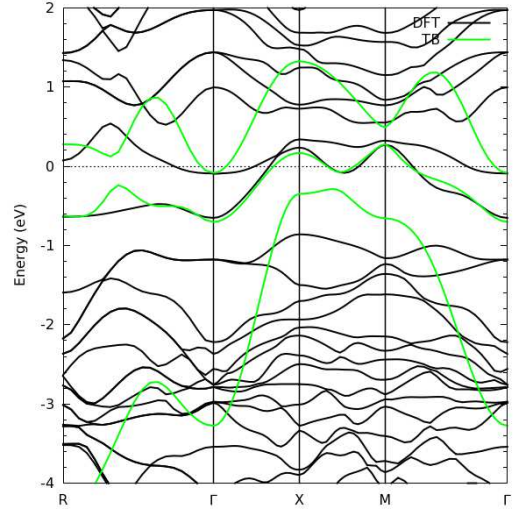}
\caption{Band structure obtained in the tight-binding model (green) is compared with that in DFT calculations (black). One can see that the essential properties around the FS are captured. DFT bands away from the FS should be related with many extra orbitals and are not attempted to be explained in this model.}\label{band}
\end{center}
\end{figure}

\begin{figure}[htb]
\begin{center}
\includegraphics[width=80mm]{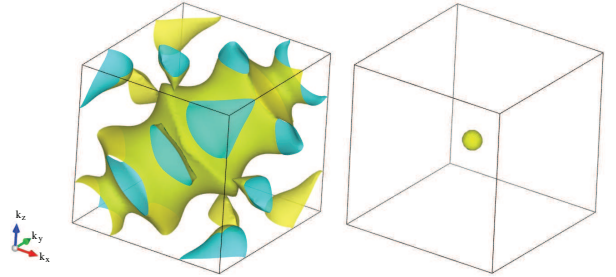}
\caption{Fermi surface of the top two bands, which agree with the DFT calculations that can be found in Ref.~\onlinecite{Cao}}\label{FS}
\end{center}
\end{figure}

{\it Model Hamiltonian.} The lattice structure of $Ir_{3}Ch_{8}$ can be found in Ref.~\onlinecite{Hosono}. DFT calculations have shown that only the locally defined $d_{z^{'2}}$ orbitals, the orientations of which are shown in Fig.~\ref{structure}, on the $Ir$ sites, and the $\sigma^{*}$ state on the $Ch_{2}$ located at the center of the unit cell appear on the FS\cite{Hosono}.  This observation allow us to build a minimum effective model.  As shown in Fig.\ref{structure},  the essential coupling related to the $Ir$ orbitals must take place around the centers of the circles denoted as $A$ and $B$ where the $Ir$ $d_{z^{'2}}$ orbitals overlap with each other and the $\sigma^{*}$ state.  Therefore, we can use the $\sigma^{*}$ state, which is denoted as $|C\rangle$, and the other two $Ir$ Wannier states $|A\rangle$ and $|B\rangle$ located around $A$ and $B$ to construct the effective band structure near the FS. Ignoring the small rhombohedral distortion\cite{Hosono, Cao}, the Hamiltonian can be written as:
\begin{equation}\begin{split}
H_{AA}&=2t_0(cos(k_x)+cos(k_y)+cos(k_z))+2t^{'}_{0}(cos(k_x+k_y+k_z))\\
      &+4t^{''}_{0}(cos(k_x-k_y)+cos(k_z-k_y)+cos(k_x-k_z))\\
      &+4t^{'''}_{0}(cos(k_x+k_y)+cos(k_z+k_y)+cos(k_x+k_z))-\mu\\
H_{BB}&=H_{AA}\\
H_{CC}&=2t_c(cos(k_x)+cos(k_y)+cos(k_z))\\
      &+2t^{'}_{c}(cos(k_x+k_y+k_z))+\Delta-\mu\\
      &+4t^{''}_{c}(cos(k_x)cos(k_y)+cos(k_z)cos(k_y)+cos(k_x)cos(k_z))\\
H_{BA}&=t_{ab}e^{i(k_x+k_y+k_z)/3}+t^{'}_{ab}e^{-2i(k_x+k_y+k_z)/3}\\
     &+t^{''}_{ab}(e^{i(-2k_x+k_y+k_z)/3}+e^{i(k_x-2k_y+k_z)/3}+e^{i(k_x+k_y-2k_z)/3})\\
H_{CA}&=t_{1}e^{-i(k_x+k_y+k_z)/6}+t_{2}e^{i5(k_x+k_y+k_z)/6}=-H_{CB}^{*}
\end{split}\end{equation}
where $\mu$ is the chemical potential and $\Delta$ is the energy difference between the $Ir$ states and the $Ch_{2}$ $\sigma^{*}$ state.
The hopping parameters (in the unit of $|t_1|$ which is around $0.3 eV$) obtained by fitting the DFT bands around the FS\cite{Hosono, Cao} are listed in Table.\ref{tab:Dab3}.
\begin{table}[H]
 \centering
 \begin{tabular}{|p{1.0cm}p{1.0cm}p{1.0cm}p{1.0cm}p{1.0cm}p{1.0cm}p{1.0cm}|}\hline

\hline
 $t_0$       &$t^{'}_{0}$       &$t^{''}_{0}$       &$t^{'''}_{0}$      &$t_{ab}$             &$t^{'}_{ab}$      &$t^{''}_{ab}$     \\
\hline
 $0.28$   &$-0.2$     &$-0.75$   &$-0.2$      &$0.4$   &$-0.3$  & $-1.8$   \\
\hline
 $t_1$     &$t_2$    &$t_{c}$    &$t^{'}_{c}$       &$t^{''}_{c}$  &$\mu$       &$\Delta$  \\
\hline
 $-1.0$    &$0.8$    &$-0.2$     &$-0.1$            & $-0.1$      &$1.1$   &$1.5$   \\
\hline
 \end{tabular}
\caption{Hopping parameters in the effective model
\label{tab:Dab3}} 
\end{table}
 The charge density given by this model is around $1.4$ electrons per unit cell per spin. The energy bands and FS in this model are shown in Fig.~\ref{band} and Fig.~\ref{FS} and it can be concluded from the figures that our effective model captures the main electronic physics near the FS. We do not try to fit anything away from the FS because it will dramatically increase the desired minimum number of bands and hence the complicity. The FS of the major band has little dispersion along the $(111)$ direction and hence quasi-$2d$, which agrees with DFT calculations~\cite{Cao}.


{\it Interactions and Loop Current.}  As stated above, the on-site interactions are suppressed in these materials. At the same time, the three $Ir$ atoms forming a state $|A\rangle$ or $|B\rangle$ have their $d_{z^{'2}}$ orbitals pointing together, which will evidently cause a repulsion between them:   
\begin{equation}\begin{split}\label{interaction}
V&=\sum_{<\mathbf{i},\mathbf{j}>}Un_{\mathbf{i}}n_{\mathbf{j}},\\
\end{split}
\end{equation} 
where $n_{\mathbf{i}}=n_{\mathbf{i}\uparrow}+n_{\mathbf{i}\downarrow}$ and $n_{\mathbf{i}\sigma}$ is the electron number operator with spin $\sigma$ in the $d_{z^{'2}}$ orbital on the $i^{th}$ $Ir$ site. 

 Diamagnetism may be generated with this interaction. To show this, we introduce the current operator
\begin{equation}\begin{split}\label{interaction} 
J_{\mathbf{i,j}\sigma\sigma^{'}}=i(c^{\dagger}_{\mathbf{i}\sigma}c_{\mathbf{j}\sigma^{'}}-H.C.) 
\end{split}
\end{equation} 
and then obtain
\begin{equation}\begin{split}\label{interaction}
n_{\mathbf{i}\sigma}n_{\mathbf{j}\sigma^{'}}&=\frac{1}{2}(-|J_{\mathbf{i,j}\sigma\sigma^{'}}|^{2}+n_{\mathbf{i}\sigma}+n_{\mathbf{j}\sigma^{'}})\\
=&-\langle J_{\mathbf{i,j}\sigma\sigma^{'}}\rangle J_{\mathbf{i,j}\sigma\sigma^{'}}+\frac{1}{2}(\langle J_{\mathbf{i,j}\sigma\sigma^{'}}\rangle^{2}+n_{\mathbf{i}\sigma}+n_{\mathbf{j}\sigma^{'}})\\
\end{split}
\end{equation} 
with a mean-field approximation. 

 For a charge loop current state we only keep the interaction between electrons with the same spin in our consideration and take the mean-field ansatz~\cite{Varma} 
\begin{equation}\begin{split}\label{interaction}
\frac{U}{2}\langle J_{\mathbf{i,j}\sigma\sigma}\rangle=-r\\
\end{split}\end{equation} 
, where $<\mathbf{i,j}>=<\mathbf{2,1}>$, $<\mathbf{3,2}>$, or $<\mathbf{1,3}>$ (see Fig.~\ref{structure}). Here a negative value of $r$ means a loop current floating along the direction $\mathbf{1}\rightarrow\mathbf{2}\rightarrow\mathbf{3}\rightarrow\mathbf{1}$, while a positive one represents a current running oppositely. 
With this mean-field ansatz, we are able to omit the spin index and write $V$, up to a constant, to be $V= \sum_{\mathbf{k}}V(\mathbf{k})$ with
\begin{equation}\begin{split}
V(\mathbf{k})&=4ri(cos\frac{k_{x}-k_{y}}{2}c^{\dagger}_{2\mathbf{k}}c_{1\mathbf{k}}
+cos\frac{k_{z}-k_{x}}{2}c^{\dagger}_{3\mathbf{k}}c_{2\mathbf{k}}\\
&+cos\frac{k_{y}-k_{z}}{2}c^{\dagger}_{1\mathbf{k}}c_{3\mathbf{k}})+H.C.+2U(n_{1\mathbf{k}}+n_{2\mathbf{k}}+n_{3\mathbf{k}}).\\
\end{split}\end{equation}

In order to study this loop current with the tight-binding model, we need to write down $V$ in the Wannier states $|\mathbf{k},A(B)\rangle$ representation. The three eigenstates of $V(\mathbf{k})$ can be solved as $|\mathbf{k},l\rangle$ with eigenvalues $E=4lrR(\mathbf{k})+2U$ where $l=0,\pm1$, and $R=\sqrt{cos^{2}\frac{k_{x}-k_{y}}{2}+cos^{2}\frac{k_{y}-k_{z}}{2}+cos^{2}\frac{k_{z}-k_{x}}{2}}$. In general, the two Wannier states $|\mathbf{k},A(B)\rangle$ can be expressed as superpositions of these three states $|\mathbf{k},l\rangle$. We notice that, independent on the value of $r$, $|\mathbf{k},0\rangle$ is always the second lowest state so that its weight should not vanish when we are going to express two states. Hence we are able to write $|\mathbf{k},A(B)\rangle$ as:   
\begin{equation}\begin{split}
|\mathbf{k},A\rangle&=\frac{|\mathbf{k},0\rangle+\alpha|\mathbf{k},1\rangle+\beta|\mathbf{k},-1\rangle}{\sqrt{1+\alpha^{2}+\beta^{2}}}\\
|\mathbf{k},B\rangle&=\frac{|\mathbf{k},0\rangle+\theta|\mathbf{k},1\rangle+\eta|\mathbf{k},-1\rangle}{\sqrt{1+\theta^{2}+\eta^{2}}},\\
\end{split}\end{equation}
where $\alpha$, $\beta$, $\theta$ and $\eta$ are coefficients and the orthonormal conditions require them to satisfy $\alpha=-\theta$, $\beta=-\eta$, $\alpha^{2}+\beta^{2}=1$.
Then the matrix elements of the interaction term in the $|\mathbf{k},A(B)\rangle$ basis can be estimated to be (applying $\delta=\alpha^{2}-\beta^{2}=2\alpha^{2}-1$)
\begin{equation}\begin{split}
V_{AA}&=V_{BB}=2U+2rR\delta,\\
V_{BA}&=V_{AB}=-2rR\delta.\\
\end{split}\end{equation}
 
Since the Wannier states depend on $\delta$, the loop current order also modifies the tight-binding part of the Hamiltonian so that there will be a competition between the hopping and the loop current. First of all, a locally rotating state does not couple with $Ch_{2}$ due to symmetry argument: the hopping parameters between a $Ch_{2}$ and the $3$ surrounding $Ir$ atoms in a rotating state must have the same amplitude and a $\pm \frac{2\pi}{3}$ phase difference. Secondly, for the hopping between the $Ir$ states, we omit the $\mathbf{k}$ index and  apply $H_{ll^{'}}=\langle l|H|l^{'}\rangle$, and notice that $|1\rangle=|-1\rangle^{*}$ and $H_{AA}=H_{BB}$ are guaranteed by symmetry, then we can use real numbers $\xi_{1}=\Re(H_{1-1})$ and $\xi_{2}=\Im(H_{01})$ to denote the change in hopping:
\begin{equation}\begin{split}
dH_{AA}&=dH_{BB}=\xi_{1}d(\alpha\beta), \\
dH_{BA}&=i\xi_{2}(d\alpha-d\beta)-\xi_{1}d(\alpha\beta). \\
\end{split}\end{equation} 

\begin{figure}[htb]
\begin{center}
\includegraphics[width=80mm]{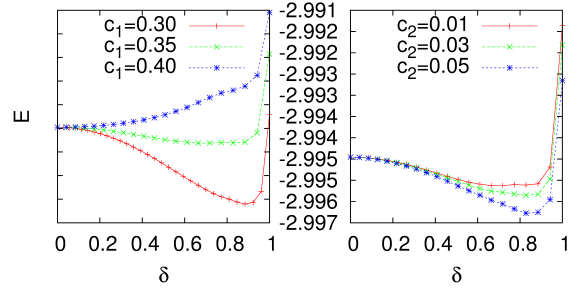}
\caption{$E$ $vs$ $\delta$ at $U=1.0$, $c_{2}=0.01$, variable $c_{1}$ (left panel) and $c_{1}=0.35$, variable $c_{2}$ (right panel).}\label{VEc12}
\end{center}
\end{figure}

For simplicity but without losing the essential physics, we assume the $\xi_{i} (i=1,2)$ parameters are proportional to the overlap of the wave functions which is measured by the original $H_{AA}|_{\alpha=\beta}$ and set $\xi_{i}=c_{i}H_{AA}|_{\alpha=\beta}$. Then the total energy of the system can be calculated self-consistently and the influence of $c_1$ and $c_{2}$ is shown in Fig.~\ref{VEc12}. It is clear that $c_{1}$ costs energy and is able to compete with $U$, while $c_{2}$ saves energy. We know that the loop current can not be a huge effect in real materials so that $c_{1}\gg c_{2}$, and hence the value of $c_{2}$ is not important as long as it is much smaller than $c_{1}$. 

\begin{figure}[htb]
\begin{center}
\includegraphics[width=80mm]{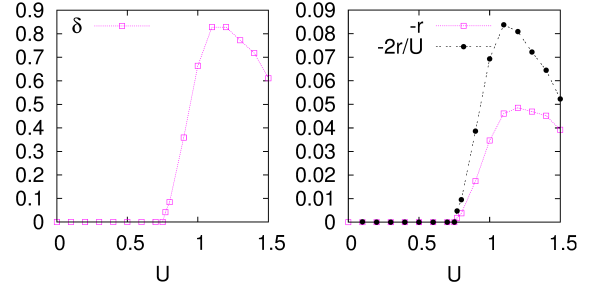}
\caption{Optimal $\delta$ $vs$ $U$ (left panel) and $-r$ and $\langle J\rangle=\frac{-2r}{U}$  $vs$ $U$ (right panel) at $c_{1}=0.35,c_{2}=0.01$.}\label{Vdr}
\end{center}
\end{figure}

To obtain the values of $r$ and $\langle J\rangle$, for each $U$  we vary the parameter $\delta$ and calculate $r$ and $E$, then choose the value of $\delta$ which gives the lowest energy as the optimal one. Fig.~\ref{Vdr} shows the relationship between optimal $\delta$, $r$, $\langle J\rangle$ and $U$, and a phase transition is clearly obtained to be around $U=0.75$. As $U$ increases, the mean value of the current $\langle J\rangle$ firstly grows up as expected, and then goes down because when $U$ is too large electrons will be moved from the $Ir$ atoms to the $Ch_{2}$ and the current is suppressed. The $c's$ are chosen so that a reasonable value of $U$ can induce a small loop current.

Hence we see that the inter-site repulsion given by Eq.~\ref{interaction} is able to induce a loop current state in which diamagnetism is certainly deserved because the loops react against any magnetic flux penetrating through them. And because the loops are perpendicular to the $(111)$ axis, the diamagnetism must be highly anisotropic and most sensitive to a magnetic field along the $(111)$ direction as observed experimentally\cite{Cao}. 

\begin{figure}[H]
\begin{center}
\includegraphics[width=40mm]{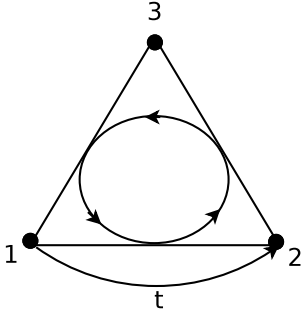}
\caption{The loop is considered as an inscribed circle of the triangle.}\label{loop}
\end{center}
\end{figure}

The amplitude of the loop magnetic moment $m$ can also be estimated in this model. Assuming the hopping between the $d_{z^2}$ orbitals of the three $Ir$ sites forming a loop is $t$, then the measured current density should be $t\langle J\rangle$. Treating the loop as an inscribed circle of the triangle made by the three $Ir$ atoms as shown in Fig.~\ref{loop}, then we have $m=I\pi R_{0}^2=-\frac{\sqrt{3}e r t a^2}{6\hbar U}$ where $R_0$ is the radius of the circle, while $e$ is the elementary charge and $a$ is the lattice constant, so that only $\frac{-rt}{U}$ is unknown. We estimate $-\frac{r}{U}\approx 0.02$ from Fig.~\ref{Vdr} and $t\approx 0.5eV$. Then one can obtain $m=0.022\mu_B/Ir$. 
This loop magnetic moment is, although not quantitatively accurate, of the same order of magnitude as the $0.024\mu_B/Ir$ paramagnetic moment obtained by fitting the experimental data~\cite{Cao}, and hence should be important in the study on the magnetic structure.

{\it Fluctuations and Pairing.} The possibility for the fluctuations of the loop current to be the origin of superconductivity has been studied in cuprates\cite{Varma2,Varma1,Ronny,Ronny1}. In the pyrite-type $Ir_{x}Ch_{2}$ iridium chalcogenide superconductors, it is observed that superconductivity is achieved or enhanced by doping more $Ir$ atoms, instead of just more charge carriers as in the high $T_{c}$ case, into the parent compound\cite{Hosono}. The stability of the loop current relies on the fact that the three $Ir$ atoms in the same loop strongly couple together, and hence extra $Ir$ sites can disturb the coupling and enhance fluctuations of the loop current. All of these imply that in these iridium chalcogenides, the fluctuations of the loop current play an important role in the superconductivity.

Upon the loop current state described above in which there is a global current order parameter $r$, we consider the fluctuation that for a loop whose center is located at $\mathbf{i}$ the current order parameter varies by $\rho_{\mathbf{i}}$: $r_{\mathbf{i}}=r+\rho_{\mathbf{i}}$.
For the dynamics of $\rho_{\mathbf{i}}$ we consider the harmonic restoration force from loop $\mathbf{i}$ itself (with spring constant $\kappa_{1}$) and the other three loops with which it shares a common $Ir$ (with spring constant $\kappa_{2}$), and seek the solution of the form $\rho_{\mathbf{i}}\sim e^{i(\mathbf{q}\cdot\mathbf{i}-\omega t)}$, then we obtain:
\begin{equation}\begin{split}
\omega(\mathbf{q})&=[\kappa_{1}-\kappa_{2}(cos\frac{-2q_{x}+q_{y}+q_{z}}{3}\\
&+cos\frac{-2q_{y}+q_{z}+q_{x}}{3}+cos\frac{-2q_{z}+q_{y}+q_{x}}{3})]^{0.5}.\\
\end{split}\end{equation}
In a perturbative approximation, the pairing strength induced by such fluctuations of the loop current should be proportional to 
\begin{equation}\begin{split}\label{pairing}
&\sum_{\mathbf{q}}\frac{\omega(\mathbf{q})}{(E(\mathbf{k+q})-E(\mathbf{k}))^2-\omega^{2}(\mathbf{q})}.\\
\end{split}\end{equation}
One can see that for any $\mathbf{Q}$ along the $(111)$ direction, $\omega(\mathbf{Q})$ has its minimal value $\sqrt{\kappa_1-3\kappa_2}$ even if the values of $\kappa_1$ and $\kappa_2$ are hard to be estimated. Moreover, for any state $|\mathbf{k} \rangle$ around the nearly cylindrical Fermi surface, we have $E(\mathbf{k+Q})\approx$ $E(\mathbf{k})\approx E_{F}$, and hence with small $\omega(\mathbf{Q})$ pairing can be induced on the entire quasi-2d Fermi surface according to Eq.\ref{pairing}.

{\it Summary and Discussions.} In this paper, we present a three-band tight-binding model to describe the electronic properties near the FS of $Ir_{3}Ch_{8}$, the parent compound of the pyrite-type $Ir_{x}Ch_{8}$ superconductors. We show the model has a loop current state as a ground state because the on-site interaction and super-exchange processes are suppressed and the inter-site repulsion emerges as the most important interaction.  The loop current  state   carries a weak magnetic moment and exhibits  a strong and highly anisotropic diamagnetism. The results are consistent  with experiments. We also suggest that the fluctuation of the loop current  can be the origin of the superconducting pairing.     

{\it Acknowldgements.} This work is supported by the Ministry of Science and Technology of China 973 program(2012CB821400).

\end{document}